%% file: ms.tex
\documentclass[12pt,article]{emulateapj} 

\shorttitle{Spatially resolved emission in a high$-z$ DLA galaxy}
\shortauthors{Jorgenson \& Wolfe}

\input{xavier_defs.tex}

\usepackage{graphicx}

\usepackage{lscape} 
\usepackage{pdflscape}
\usepackage{natbib}
\begin{document}

\def\intl{\int\limits}
\def\nstat{$\approx $}
\def\perd{\;\;\; .}
\def\cmma{\;\;\; ,}
\def\ltk{\left [ \,}
\def\ltp{\left ( \,}
\def\ltb{\left \{ \,}
\def\rtk{\, \right  ] }
\def\rtp{\, \right  ) }
\def\rtb{\, \right \} }
\def\jnu{$J_{\nu}$}
\def\jnuphot{$J_{\nu}^{phot}$}
\def\jnuciistar{$J_{\nu}^{\rm CII^{*}}$}
\def\junit{ergs cm$^{-2}$ s$^{-1}$ Hz$^{-1}$ sr$^{-1}$}
\def\jnutot{$J_{\nu}$$^{total}$}
\def\jnubkd{$J_{\nu}$$^{Bkd}$}
\def\jnuloc{$J_{\nu}$$^{local}$}
\def\jnulw{$J_{\nu}$$^{LW}$}
\def\jnutotciistr{$J_{\nu }$$^{total, C\,II^*}$}
\def\jnulocciistr{$J_{\nu } $$^{local, C\,II^*}$}
\def\jnutotci{$J_{\nu }$$^{total, C\, I}$}
\def\jnulocci{$J_{\nu }$$^{local, C\, I}$}
\def\jnutothtwo{$J_{\nu }$$^{total, H_2}$}
\def\jnulochtwo{$J_{\nu }$$^{local, H_2}$}
\newcommand{\snrlim}{SNR$_{lim}$}
\newcommand{\nhi}{$N_{\rm HI}$}
\newcommand{\mnhi}{N_{\rm HI}}
\newcommand{\flls}{f_{\rm HI}^{\rm LLS}}
\newcommand{\fdla}{f_{\rm HI}^{\rm DLA}}
\newcommand{\llls}{$\ell_{\rm LLS}$}
\newcommand{\ldla}{\ell_{\rm DLA}}
\newcommand{\fnhi}{$f_{\rm HI}(N,X)$}
\newcommand{\mfnhi}{f_{\rm HI}(N,X)}
\newcommand{\Nth}{2 \sci{20} \cm{-2}}
\newcommand{\taux}{$d\tau/dX$}
\newcommand{\gz}{$g(z)$}
\newcommand{\nz}{$n(z)$}
\newcommand{\nx}{$n(X)$}
\newcommand{\omg}{$\Omega_g$}
\newcommand{\ostr}{$\Omega_*$}
\newcommand{\momg}{\Omega_g}
\newcommand{\olls}{$\Omega_g^{\rm LLS}$}
\newcommand{\odla}{$\Omega_g^{\rm DLA}$}
\newcommand{\oneut}{$\Omega_g^{\rm Neut}$}	
\newcommand{\ohi}{$\Omega_g^{\rm HI}$}
\newcommand{\olwz}{$\Omega_g^{\rm 21cm}$}
\newcommand{\ndla}{71}
\newcommand{\cmk}{cm$^{-3}$ K }
\newcommand{\ci}{C\,I}
\newcommand{\cistr}{C\,I$^{*}$}
\newcommand{\mcistr}{C\,I^{*}}
\newcommand{\cistrstr}{C\,I$^{**}$}
\newcommand{\mcistrstr}{C\,I^{**}}
\newcommand{\citot}{(C\,I)$_{tot}$}
\newcommand{\mcitot}{(C\,I)_{tot}}
\newcommand{\cli}{Cl\,I}
\newcommand{\clii}{Cl\,II}
\newcommand{\cii}{C\,II}
\newcommand{\ciistr}{C\,II$^*$}
\newcommand{\dla}{DLA}
\newcommand{\dlas}{DLAs}
\newcommand{\htwo}{H$_{\rm 2}$}
\newcommand{\he}{He\, I}
\newcommand{\sii}{Si\,II}
\newcommand{\siistr}{Si\,II$^{*}$}
\newcommand{\hi}{H\, I}
\newcommand{\ctwo}{C\,II}
\newcommand{\ewsitwo}{$W_{\lambda 1526}$}
\newcommand{\ewciv}{$W_{\lambda 1548}$}
\newcommand{\feii}{Fe\,II}

\newcommand{\halpha}{H$\alpha $}
\newcommand{\hamath}{H\alpha }
\newcommand{\oiii}{[O\,III]}
\newcommand{\nii}{[N\,II]}

\newcommand{\mydla}{\dla 2222$-$0946}

\newcommand{\zfynbo}{$z = 2.35409$}
\newcommand{\zfynbooiii}{$z = 2.35406$}
\newcommand{\zfynboblue}{$z = 2.35341$}
\newcommand{\zfynbored}{$z = 2.35445$}

\newcommand{\zkrog}{$z=2.3537}

\newcommand{\hazabs}{2.35391}
\newcommand{\havel}{$-$44}
\newcommand{\haflux}{(4.76 $\pm$ 0.50) $\times 10^{-17}$}
\newcommand{\halum}{(2.13 $\pm$ 0.23) $\times 10^{42}$}
\newcommand{\hasfr}{9.5 $\pm$ 1.0}
\newcommand{\hafwhm}{144}
\newcommand{\hafwhminit}{118}
\newcommand{\hasig}{50}
\newcommand{\hamass}{6.1 $\times$ 10$^9$}
\newcommand{\niizabs}{2.35361}
\newcommand{\niivel}{$-$71}
\newcommand{\niiflux}{(7.83 $\pm$ 4.69) $\times 10^{-18}$}
\newcommand{\niilum}{(3.51 $\pm$ 2.1) $\times 10^{41}$}
\newcommand{\niifwhm}{113}
\newcommand{\niimetN}{8.45 $\pm$ 0.26}
\newcommand{\niimetsolar}{0.62}
\newcommand{\niimetsolarpercent}{62\%}

\newcommand{\niizabsopt}{2.35384}
\newcommand{\niivelopt}{$-$50}
\newcommand{\niifluxopt}{(1.48 $\pm$ 0.46) $\times 10^{-18}$}
\newcommand{\niilumopt}{(6.62 $\pm$ 2.1) $\times 10^{40}$}
\newcommand{\niifwhmopt}{124}
\newcommand{\niifwhmoptinit}{92}
\newcommand{\niimetNopt}{8.54 $\pm$ 0.14}
\newcommand{\niimetsolaropt}{0.75}
\newcommand{\niimetsolaroptpercent}{75\%}
\newcommand{\niimetNoptplusdispersion}{8.54 $\pm$ 0.32}
\newcommand{\hafluxopt}{(6.38 $\pm$ 0.66) $\times 10^{-18}$}
\newcommand{\hazabsopt}{2.35386}
\newcommand{\havelopt}{$-$48}

\newcommand{\sigsfr}{0.55}
\newcommand{\sigsfrpeak}{1.7}
\newcommand{\siggas}{243}
\newcommand{\totgas}{4.2 $\times$ 10$^9$}

\newcommand{\oiiiazabs}{2.35397}
\newcommand{\oiiiavel}{$-$39}
\newcommand{\oiiiaflux}{(7.87 $\pm$ 0.62) $\times 10^{-17}$}
\newcommand{\oiiialum}{(3.52 $\pm$ 0.28) $\times 10^{42}$}
\newcommand{\oiiiafwhm}{155}
\newcommand{\oiiiafwhminit}{131}
\newcommand{\oiiibzabs}{2.35406}
\newcommand{\oiiibvel}{$-$30}
\newcommand{\oiiibflux}{(2.91 $\pm$ 0.74) $\times 10^{-17}$}
\newcommand{\oiiiblum}{(1.31 $\pm$ 0.33) $\times 10^{42}$}
\newcommand{\oiiibfwhm}{138}
\newcommand{\oiiibfwhminit}{111}
\newcommand{\oiiiratio}{2.7}

\newcommand{\hireszabs}{2.35440}
\newcommand{\hireszabsblue}{2.3530} 
\newcommand{\hireszabsred}{2.3563}

\newcommand{\jcap}{Journal of Cosmology and Astroparticle Physics}

\def\lc{${\ell}_{c}$}
\def\lcunit{ergs s$^{-1}$ H$^{-1}$}
\newcommand{\delvninty}{$\Delta v_{90}$}

\title{Spatially Resolved Emission of a High Redshift DLA Galaxy with the Keck/OSIRIS IFU\altaffilmark{1}}

\author{Regina A. Jorgenson\altaffilmark{2} \& Arthur M. Wolfe\altaffilmark{3}
}

\altaffiltext{1}{The data presented herein were obtained at the W.M. Keck Observatory, which is operated as a scientific partnership among the California Institute of Technology, the University of California and the National Aeronautics and Space Administration. The Observatory was made possible by the generous financial support of the W.M. Keck Foundation. }

\altaffiltext{2}{Institute for Astronomy, University of Hawaii, 2680 Woodlawn Drive, Honolulu, HI, 96822; raj@ifa.hawaii.edu}

\altaffiltext{3}{Department of Physics and Center for Astrophysics and Space Sciences, University of California, San Diego, 9500 Gilman Dr., La Jolla, CA 92093-0424, USA}

\begin{abstract}

We present the first Keck/OSIRIS infrared IFU observations of a high redshift damped Lyman$-\alpha$ (\dla ) galaxy detected in the line of sight to a background quasar.  By utilizing the Laser Guide Star Adaptive Optics (LGSAO) to reduce the quasar PSF to FWHM$\sim$ 0.15$\arcsec$, we were able to search for and map the foreground \dla\ emission free from the quasar contamination.  We present maps of the \ha\ and \oiii\ $\lambda \lambda$5007, 4959 emission of \mydla\ at a redshift of $z \sim$ 2.35.  From the composite spectrum over the \ha\ emission region we measure a star formation rate of \hasfr\ M$_{\odot}$ year$^{-1}$ and a dynamical mass, M$_{dyn}$ = \hamass\ M$_{\odot}$.  The average star formation rate surface density is $\langle \Sigma _{SFR} \rangle$ = \sigsfr\ M$_{\odot}$ yr$^{-1}$ kpc$^{-2}$, with a central peak of \sigsfrpeak\ M$_{\odot}$ yr$^{-1}$ kpc$^{-2}$.  Using the standard Kennicutt-Schmidt relation, this corresponds to a gas mass surface density of $\Sigma _{gas}$ = \siggas\ M$_{\odot}$ pc$^{-2}$.  Integrating over the size of the galaxy we find a total gas mass of M$_{gas}$ = \totgas\ M$_{\odot}$.  We estimate the gas fraction of \mydla\ to be $f_{gas} \sim 40\%$.  We detect \nii $\lambda$6583 emission at 2.5$\sigma$ significance with a flux corresponding to a metallicity of \niimetsolaroptpercent\ solar.  Comparing this metallicity with that derived from the low-ion absorption gas $\sim$6 kpc away, $\sim$30\% solar, indicates possible evidence for a metallicity gradient or enriched in/outflow of gas.  Kinematically, both \ha\ and \oiii\  emission show relatively constant velocity fields over the central galactic region.  While we detect some red and blueshifted clumps of emission, they do not correspond with rotational signatures that support an edge-on disk interpretation. 

\end{abstract}


\section{Introduction}

The emerging picture of galaxy formation and evolution at high redshift is currently dominated by observations of the star formation rate per unit comoving volume from $z \gtrsim\ $7 to the present day and shows that 50\% of the current stellar mass of galaxies formed in the redshift interval 1.5 $< z < $ 3.5 ~\citep{reddy09}.  Photometric surveys for galaxies have succeeded in tracing their stellar content out to redshifts as large as 6 or higher  \citep{ellis13, lehnert10, 2004ApJ...600L.103G, 2004ApJ...606L..25B}.  The majority of galaxies found in this way are the Lyman Break Galaxies (LBGs; e.g. \cite{steidel03}), which are selected for bright rest-frame UV emission.  These are star-forming (mean SFR $\sim$ 40 M$_{\odot}$ yr$^{-1}$) galaxies with  average half-light radii $<r_{1/2}>$ $\approx$ 2 kpc \citep{2004ApJ...612..108S}.  Because  
they are strongly clustered ($r_{0}$$\approx$ 4$h^{-1}$ Mpc; \cite{2005ApJ...620L..75A}), the LBGs are likely to be biased tracers of dark-matter halos with masses, $M_{DM}$$\sim$10$^{+12}$ M$_{\odot}$. Consequently, the LBGs were originally thought to be the progenitors of massive elliptical galaxies \citep{1999ApJ...519....1S, 2003ApJ...592..728S}. However, recent studies of \halpha\  emission at $z$ $\approx$ 2.5 with the SINFONI IFU on the VLT and OSIRIS on Keck suggest some fraction could be the progenitors of massive spiral galaxies.  This follows from the detection of disks rotating with circular velocities $v_{c}$ $\approx$ 200 to 250 {\kms} \citep{law12, 2009ApJ...706.1364F, genzel06}. 

The possibility that massive spirals were in place at $z$ $>$ 2 has important implications for hierarchical theories of galaxy formation, which predict most objects at $z$ $>$ 2 to have $v_{c}$ $<<$ 250 {\kms}. Therefore, it is crucial to determine whether ``rotating'' LBGs are representative protogalaxies or rarely occurring luminous ``5-$\sigma$ events."  In addition, the link between massive, high star formation rate LBGs, and the neutral atomic gas that must be the fuel for their copious star formation remains unclear.  Recent models have connected the Damped Lyman alpha absorption systems (DLAs), another class of high-$z$ objects that qualify as spiral progenitors, with the LBGs ~\citep{wolfe08}.  The DLAs are drawn from  a cross-section weighted sample of neutral gas layers, which contain sufficient neutral gas to account for most of the visible stars in modern galaxies, and with properties resembling those of spiral disks \citep{wolfe05}. Interestingly, the DLA absorption-line kinematics  are consistent with randomly oriented disks with $v_{c}$ $\approx$ 250 {\kms}. But since the velocity fields are deduced from absorption-line studies alone, the \dla\ masses and sizes are generally unknown.  However, ~\cite{cooke05, cooke06} cross-correlate \dlas\ with LBGs at $z \sim$ 3 and find that they reside in similar spatial locations and have a similar inferred dark matter halo mass range of 10$^{9-12}$M$_{\odot}$. Alternate models have suggested that \dla\ velocity profiles are consistent with merging protogalactic clumps of gas predicted by SPH simulations of structure formation (~\cite{haehnelt98, hong10}, however see ~\cite{pro10} who point out several mistreatments in these works).   

Despite evidence for star formation ~\citep{wolfe03a} and metal enrichment ~\citep{rafelski12, jorgenson13a} in \dlas , the direct detection of \dlas\ in emission has been rare.  Efforts to image \dlas\ directly have generally been unsuccessful because of the difficulty of detecting relatively faint foreground emission near a much brighter background quasar (i.e. ~\cite{lowenthal95, bunker99, kulkarni00, kulkarni06, christensen09} and several unpublished works).  To date, only ten $z > $ 2 \dlas\ have been detected in emission (see ~\cite{krogager12} for a summary).  All of these targets, with the exception of one, \mydla , discussed in this paper, were detected in single slit observations, requiring fortuitous slit placement and providing limited information on the total fluxes, star formation rates (SFR) and kinematics of the emission.  

The advent of Laser Guide Star Adaptive Optics (LGSAO) corrected Integral Field Units (IFU) on 10-meter class telescopes such as the Keck/OSIRIS IFU ~\citep{larkin06}, creates a clear, new path forward to answering some questions raised since the first surveys of \dlas ~\citep{wolfe86}.  By taking advantage of the LGSAO correction to minimize the PSF of the background quasar, it is possible to search for \dla\ emission at small impact parameters while simultaneously obtaining spectra that provide mass and kinematic measures. Will a majority of high$-z$ \dlas\ reveal disk-like rotation, further challenging the hierarchical theory of structure formation?  
Are star formation rates as estimated by the \ciistr\ technique ~\citep{wolfe03a} and implying that $\sim$50\% of \dlas\ should be associated with the halos of LBGs ~\citep{wolfe08} correct? Only by complementing the wealth of \dla\ absorption-line data with the direct detection and mapping of emission can the true nature of these enigmatic systems be understood.

We have used the Keck/OSIRIS IFU with LGSAO to target the high metallicity \dla , \mydla , first found by ~\cite{fynbo10} to have relatively strong \lya , \ha , and \oiii $\lambda$4959, $\lambda $5007 emission in single-slit VLT/X-Shooter observations.  At a redshift of $z \sim 2.35$, \mydla\ has a neutral hydrogen gas column density of \nhi\ = 4.5 $\times$ 10$^{20}$ cm$^{-2}$, a metallicity of [M/H]\footnote{We use the standard shorthand notation for metallicity relative to solar, [M/H] = log(M/H) $-$ log(M/H)$_{\odot }$.} $\sim -0.5$ ~\citep{krogager13, jorgenson13a} and lies along the line of sight to background quasar SDSS J222256.11$-$094636.2.  

\mydla\ was imaged with the VLT/SINFONI IFU by ~\cite{peroux12} and then again by ~\cite{peroux13}, however in both cases only Natural Guide Star Adaptive Optics (NGSAO) was used, leading to point spread functions (PSF) of FWHM = 0.6$\arcsec$ and FWHM = 0.4$\arcsec$, respectively.  At the redshift of the \dla\ this PSF corresponds to $\sim$4 kpc, which could very easily mask  kinematic signatures in a compact galaxy (e.g. ~\cite{newman13}). In addition, ~\cite{peroux12} and ~\cite{peroux13} only targeted \halpha\ emission, while the \oiii\ flux is measured to be stronger from single slit observations ~\citep{fynbo10}. ~\cite{krogager13} used the {\it Hubble Space Telescope (HST)} to image the stellar continuum of \mydla\ in the rest frame optical-UV regime and estimate the SFR, stellar and dynamical masses and morphology.  From the {\it HST} imaging ~\cite{krogager13} conclude that the galaxy has a compact yet elongated morphology indicative of a galactic disk viewed edge-on.  

In this paper we present the first observations of a high redshift \dla\ in emission utilizing the Keck/OSIRIS IFU and LGSAO. We detect and map the flux and velocity field of \mydla\  in {\it both} \halpha\ and \oiii\ emission with a PSF of FWHM$\sim$0.15$\arcsec$.  While we find interesting morphological and kinematical signatures, we find no evidence of ordered edge-on disk rotation.  

The paper is organized as follows:  We describe our observations and data reduction process in Section~\ref{sec:obs}.  In Section~\ref{sec:analysis} we discuss the details of the analysis of the final data cube.  We attempt to place these results in a larger context in Section~\ref{sec:results}, before summarizing in Section~\ref{sec:summary}.  Throughout the paper we assume a standard lambda cold dark matter ($\Lambda$CDM) cosmology based on the final nine-year {\it WMAP} results ~\citep{wmap} in which $H_0$ = 70.0 km s$^{-1} \mathrm{Mpc}^{-1}$, $\Omega _{m}$ = 0.279 and $\Omega _{\Lambda}$ = 0.721.

mass, size and host galaxy.  While there is much evidence to suggest that \dlas\ are the progenitors of massive spiral galaxies, competing interpretations model them as merging clumps of gas.  

\section{Observations}~\label{sec:obs}
   
   Observations were performed using the OSIRIS ~\citep{larkin06} infrared integral-field spectrograph in combination with the Keck I LGSAO system during two half-nights on 2012 July 20 and 21.  We utilized two narrow band filters, Hn4 and Kn3, corresponding to the redshifted wavelengths of \oiii\ and \ha\ emission from \dla\ 2222$-$09, respectively.  In order to achieve the best compromise between maximizing the field of view and spatial resolution, we chose the 50 mas plate scale, which provides a field of view of 2.1$\arcsec \times 3.2\arcsec $ and 2.4$\arcsec \times 3.2\arcsec $ for the Hn4 and Kn3 filters, respectively.  The spectral resolution varies from spatial pixel to spatial pixel (spaxel), but is approximately R$\sim$3600, as confirmed by measures of the average FWHM of a series of OH-sky lines.
   
 Exposure times were 900 seconds.  In order to maximize the on-source exposure time, rather than obtaining an off-source skyframe, we utilized an A-B observing pattern in which the target was placed in the top half of the field of view for frame A and then shifted to the bottom half of the field of view for frame B.  We then used observation B as the sky frame for observation A and vice versa.  In this way we cut our field of view in half, but doubled our on-source observing time.  This procedure ensured the maximum probability for detecting faint extended emission and was particularly useful given that we already had an idea of the position of the emission from previous works ~\citep{fynbo10, peroux12, krogager13}.  A suitable, bright (R$<$17) star within 50\arcsec\ of the target was used for tip tilt correction.  All observations were taken in clear weather with good, $\sim$0.6 \arcsec\ or better, seeing.
   
   Over the course of 2 half nights we obtained 14 $\times$ 900s in the Hn4 band and 12 $\times$ 900s in the Kn3 band for a total exposure time of 3.5 hours and 3 hours for the \oiii\ and \ha\ emission, respectively.  
      
   \subsection{Data Reduction and Flux Calibration}~\label{sec:datareduction}
   
 Data reduction was performed using a combination of the Keck/OSIRIS data reduction pipeline (DRP) and custom IDL routines in a process similar to that outlined in ~\cite{law07} and ~\cite{law09}.  We used the DRP to perform the standard reduction and extraction of the three-dimensional data cubes, including the {\tt Scaled Sky Subtraction} Module.   In order to mitigate the effects of highly variable sky lines, we performed a second-pass sky subtraction using custom IDL routines to calculate the median pixel value in each spectral channel and subtract this value from all pixels within the channel to ensure a zero-flux median in all spectral slices throughout the data cube.  
 Because the galaxy and background quasar are in each frame as both a positive and a negative, they contribute net-zero to the median flux.  
  
 We then applied a telluric correction to each science frame, using the telluric standard star taken closest in time to the science frame. 
 In order to flux calibrate the data, we used the telluric standard star and a reference Vega spectrum.  As discussed by ~\cite{law09}, the uncertainty in the absolute flux calibration of LGSAO data is estimated to be $\sim 30\,\%$ due primarily to rapid and potentially significant fluctuations of the AO-corrected core of the PSF.   
 
In order to produce a final data cube we took advantage of the fact that each frame contains the relatively bright image of the quasar, and we used the peak emission of the quasar to align the frames for mosaicing.  We produced the final data cube using the OSIRIS module {\tt Mosaic Frames} and an input file containing the frame offsets determined from the quasar centroid in each frame. The frames were combined using the sigma-clipping average routine, {\tt meanclip}.  Given our observing method of keeping the quasar and \dla\ in every frame, the final data cube consists of a positive central region with negative regions above and below.  In the final analysis we considered only spaxels located in the central positive region.  
    
 Gaussian fits to the point spread function of the quasar in the mosaiced image yield FWHM $\sim$ 3.0 and FWHM $\sim$ 3.4 pixels for the K and H bands, respectively.  For the 50 mas pixel scale, this corresponds to a FWHM of $\sim$0.15\arcsec\ and $\sim$0.17\arcsec , respectively.  In order to enhance the detail of the relatively faint emission, we spatially resampled the data to 0.025\arcsec\ pixel$^{-1}$ and then smoothed with a Gaussian kernel with FWHM=0.15\arcsec , similar to that of the LGSAO PSF.  
 
 Finally, we used the IRAF package {\tt rvcorrect} to correct the spectra for the heliocentric motion of the earth such that all wavelengths and redshifts are reported in the heliocentric vacuum frame.  Redshifts were determined using the rest-frame vacuum wavelengths of H$\alpha$, \oiii , and \nii\ (i.e. 6564.614 \AA , 5008.239 \AA , and 6585.27 \AA , respectively).  We checked the spectral resolution by measuring the FWHM of OH-skylines in the vicinity of the redshifted \ha\ and \oiii\ emission-lines and confirmed that the average resolution corresponds to R $\sim$ 3600, or FWHM $\sim$ 83   km s$^{-1}$.  The reported velocity dispersions have been corrected for the instrumental FWHM (FWHM$_{instrumental}$ $\sim$ 83 \kms ) by subtracting it in quadrature. %
 
\section{Analysis}~\label{sec:analysis}
 
 In this section we describe our analysis of the final OSIRIS data cube, including the derived flux and star formation rate estimates.  
 
 \subsection{H$\alpha$ Flux, Luminosity and Star Formation Rate}

\begin{figure}
\plotone{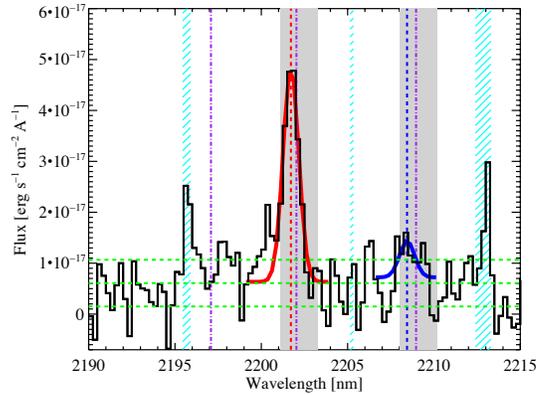}
\caption{Composite spectrum over an $\sim$0.5 arcsec$^2$ region centered on the peak of \ha\ emission.  The \ha\ emission is shown here with a best-fit Gaussian overlaid in red, while the best-fit Gaussian to the [N\, II]$\lambda$6583 is shown in blue.  The best-fit redshift of the \ha\ emission is $z = $ \hazabs , indicated here by the vertical red dashed line.  This can be compared with the fiducial absorption-line redshift, $z_{HIRES} =$ \hireszabs , indicated here by the purple dash-dot lines.  The gray vertical regions indicate the entire redshift range of the low-ion absorption.  Cyan hashed regions indicate the locations of strong atmospheric OH emission features which can leave residuals.  
}
\label{fig:halpha}
\end{figure}

 We estimate the total detected \ha\ flux, F$^{H\alpha }$, by summing the spectra in all spaxels in a $\sim 0.5\arcsec \times 0.5\arcsec $ box centered on the location of the peak \ha\ emission.  This composite spectrum is presented in Figure~\ref{fig:halpha}.  Fitting a Gaussian to the emission-line provides a  total \ha\ flux of F$^{H\alpha}$ = \haflux\ ergs s$^{-1}$ cm$^{-2}$, where the error is taken to be the standard deviation in the residual spectrum after subtraction of the Gaussian model. This corresponds to an \ha\ luminosity of  $L^{\hamath}$ = \halum\ ergs s$^{-1}$, not corrected for dust.  As in ~\cite{law09}, we use the ~\cite{kennicutt94} calibration to convert \ha\ luminosity to star formation rate (SFR),  

\begin{equation}
\mathrm{SFR (M_{\odot} \ year^{-1})} = \frac{L^{\hamath}}{1.26 \times 10^{41} \  \mathrm{erg \ s^{-1}}} \times 0.56
\label{eqn:sfr}
\end{equation}

\noindent assuming a ~\cite{chabrier03} initial mass function.  We estimate a SFR $\approx$ \hasfr\ M$_{\odot}$ year$^{-1}$.  This result is consistent, to within errors, of several previous estimates by ~\cite{fynbo10}, ~\cite{peroux12}, and ~\cite{krogager13}.  We summarize the H$\alpha$ measurements in Table~\ref{tab:results} to aid in comparison with previous works because all authors used slightly different cosmologies and assumptions to convert H$\alpha$ flux to SFR. 

\input{compare_results.tex}

From the best-fit Gaussian we estimate a FWHM(\ha ) = \hafwhm\  km s$^{-1}$, or \hafwhminit\  km s$^{-1}$, with the effects of instrumental smoothing (FWHM$_{instrumental}$ = 83 \kms ) taken out in quadrature.  We also estimate the best-fit redshift of the \ha\ emission-line to be $z = $\hazabs .  We note that this redshift is \havel\ \kms\ from $z_{HIRES} = $ \hireszabs , the redshift of a central, low-ion velocity component with the largest optical depth as measured from an archival Keck/HIRES spectrum. For the purposes of this paper, we will arbitrary define $z_{HIRES}$ to be the fiducial redshift of the system (see Section~\ref{sec:abs} for details).  This fairly large and complex absorption-line system contains several velocity components spanning the redshift range $z = $ \hireszabsblue\ to \hireszabsred .  The centroid of H$\alpha$ emission falls roughly in the middle of the absorption-line profile which is indicated by the shaded grey region in Figure~\ref{fig:halpha}. We delay a more detailed comparison between the absorption and emission-line kinematics until Section~\ref{sec:abs}. In Table~\ref{tab:summarylines} we provide a summary of all line diagnostics while in Table~\ref{tab:summaryresults} we summarize the general results.

\subsection{\nii\ }

\begin{figure}
\plotone{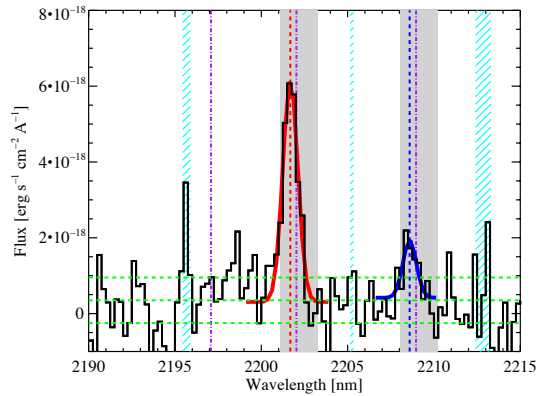}
\caption{Composite spectrum over the central $\sim$0.15 arcsec$^2$ region centered on the peak of \nii\ emission.  The \nii\ $\lambda$6583 emission-line, fit by a Gaussian in blue, is detected at 2.5$\sigma$ significance. 
The security of the line identification is raised by the fact that the independently fit \ha\ line agrees in redshift with a difference of $\sim$2  km s$^{-1}$, well within the errors.  
}
\label{fig:nii}
\end{figure}

We report a detection of the \nii $\lambda$6583 emission-line with a significance of 2.5$\sigma$.  While we attempt to fit the \nii\ emission-line with a Gaussian, shown in blue in Figure~\ref{fig:halpha}, we find that the \nii\ emission is much more centrally confined than the stronger \ha\ emission.  Therefore, to achieve the most significant detection, we created a spectral stack over only the central 0.15 arcsec$^2$ where \nii\ emission is the strongest, shown in Figure~\ref{fig:nii}.  
Line fit details are given in Table~\ref{tab:summarylines}. The best-fit Gaussian flux measures F$^{[N\,II]}$ = \niifluxopt\ ergs s$^{-1}$ cm$^{-2}$ corresponding to a luminosity $L^{[N\,II]}$ = \niilumopt\ ergs s$^{-1}$.  Under the assumption that this is a detection, we can use the $N2$ index, where $N2$ is the ratio of \nii\ to H$\alpha$ flux, to estimate the metallicity.  We use the \ha\ flux over this same region, as fit by a Gaussian (red in Figure~\ref{fig:nii}), F$^{H\alpha }$ = \hafluxopt\ ergs s$^{-1}$ cm$^{-2}$ .  As calibrated by ~\cite{pettini04} we use the relation 12 $+$ log(O/H) = 8.90 + 0.57 $\times$ $N2$ to infer a metallicity of 12 + log(O/H) = \niimetNopt .  Assuming a solar oxygen abundance of 12 $+$ log (O/H) = 8.66 ~\citep{asplund04}, the derived metallicity corresponds to \niimetsolaroptpercent\ solar metallicity.  If we include the 1$\sigma$ dispersion in the $N2$ relation, 0.18 dex, we find 12 $+$ log(O/H) = \niimetNoptplusdispersion .  The lower range, at $\sim$40\% solar, is just consistent 
with the metallicity as measured from the absorption-lines, [M/H] = $-$0.56 $\pm$ 0.1, or $\sim$30\% solar ~\citep{jorgenson13a, krogager13, fynbo10}.  We discuss the possibility of a metallicity gradient or the in/outflow of metal-enriched gas in Section~\ref{sec:nii}. We also note that over this central 0.15 arcsec$^2$ region, the independently fit redshifts of the \nii\ and \ha\ emission agree to within $\sim$2  km s$^{-1}$, as seen in Figure~\ref{fig:nii} (i.e. $z_{abs}^{\nii }$ = \niizabsopt\ and $z_{abs}^{H\alpha }$ = \hazabsopt ). 

For completeness we provide the results from fitting the \nii\ emission-line taken over the entire 0.5 arcsec$^2$ region shown in Figure ~\ref{fig:halpha}. 
While this line is only significant at the 1.5$\sigma$ level, the best-fit Gaussian flux measures F$^{[N\,II]}$ = \niiflux\ ergs s$^{-1}$ cm$^{-2}$ corresponding to a luminosity $L^{[N\,II]}$ = \niilum\ ergs s$^{-1}$.  Using the ~\cite{pettini04} calibrated relation we find a metallicity of 12 + log(O/H) = \niimetN , corresponding to \niimetsolarpercent\ solar metallicity.

 \subsection{[OIII] Flux and Luminosity}
 
\begin{figure}
\plotone{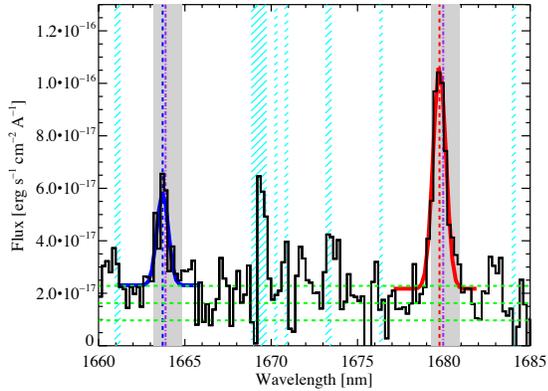}
\caption{Composite spectrum over an $\sim$0.75 arcsec$^2$ region centered on the peak of \oiii\ emission. The \oiii\ $\lambda$5007 emission is shown here with the best-fit Gaussian overlaid in red, while the blue fit on the left corresponds to the detection of \oiii $\lambda$4959.  The redshift of the \oiii $\lambda$5007 line is $z = $ \oiiiazabs\ (red dashed line), which is \oiiiavel\ km s$^{-1}$ from the fiducial absorption-line redshift, $z_{HIRES} =$ \hireszabs , indicated here by the purple dash-dot lines. The gray vertical regions indicate the entire redshift range of the low-ion absorption.  Cyan hashed regions indicate the locations of strong atmospheric OH emission features which can leave residuals.  
  }
\label{fig:oiii}
\end{figure}

 We estimate the total detected \oiii\ flux by summing the spectra in all spaxels in a $\sim$ 0.75 \arcsec\ $\times$ 0.75 \arcsec\ region around the peak \oiii\ emission.  Note that the \oiii\ emission is stronger and more spatially extended than the \halpha\ emission. We detect both \oiii\ $\lambda$5007 and \oiii\ $\lambda$4959 with high significance, as shown in Figure~\ref{fig:oiii}, where we present the composite spectrum.  Fitting a Gaussian to the \oiii\ $\lambda$5007 emission (shown in Figure~\ref{fig:oiii}, red) we measure a total \oiii\ flux of F$^{[O\,III]\lambda 5007}$ = \oiiiaflux\ ergs s$^{-1}$ cm$^{-2}$. This corresponds to an \oiii\ luminosity of  $L^{[O\,III]}$ = \oiiialum\ ergs s$^{-1}$. 

We applied an independent Gaussian fit to the \oiii\ $\lambda$4959 line and measure a total flux, F$^{[O\,III]\lambda\ 4959}$ = \oiiibflux\ ergs s$^{-1}$ cm$^{-2}$. This corresponds to an \oiii\ luminosity of  $L^{[O\,III]}$ = \oiiiblum\ ergs s$^{-1}$.  The flux ratio of F(\oiii $\lambda$5007)/F(\oiii $\lambda$4959) $\approx$ \oiiiratio , a slight deviation from the expected \oiii $\lambda$5007:\oiii $\lambda$4959 = 3:1, is perhaps insignificant given the errors in flux determination. We provide all line-fit details in Table~\ref{tab:summarylines}.  

\input{summarylines.tex}

\subsection{ Spatial mapping of intensity, velocity, velocity dispersion and signal-to-noise ratio }

 In order to map the location of emission and find kinematical signatures, we searched for emission in the spectrum of each spaxel.  For each emission-line considered, the fitting method attempted to fit a Gaussian at the expected location of emission and compared the chi-squared result to that of a fit with no emission-line.  A detection required a minimum of 4$\sigma$ to be accepted as a detection.  In this way we created a 2 dimensional map of emission for each line, where each spaxel contains a best-fit flux, velocity (relative to the best-fit redshift determined from the composite spectral stack), and velocity dispersion. The reported velocity dispersions have been corrected for the instrumental resolution by subtraction of the instrumental sigma, $\sigma _{instrumental}$ $\sim$ 35  km s$^{-1}$, in quadrature.   In cases where there was no line detected, or the signal-to-noise ratio ( S/N) was too low we do not report a detection and leave the spaxel black in the final maps.  

\begin{figure*}
\includegraphics[scale=0.78]{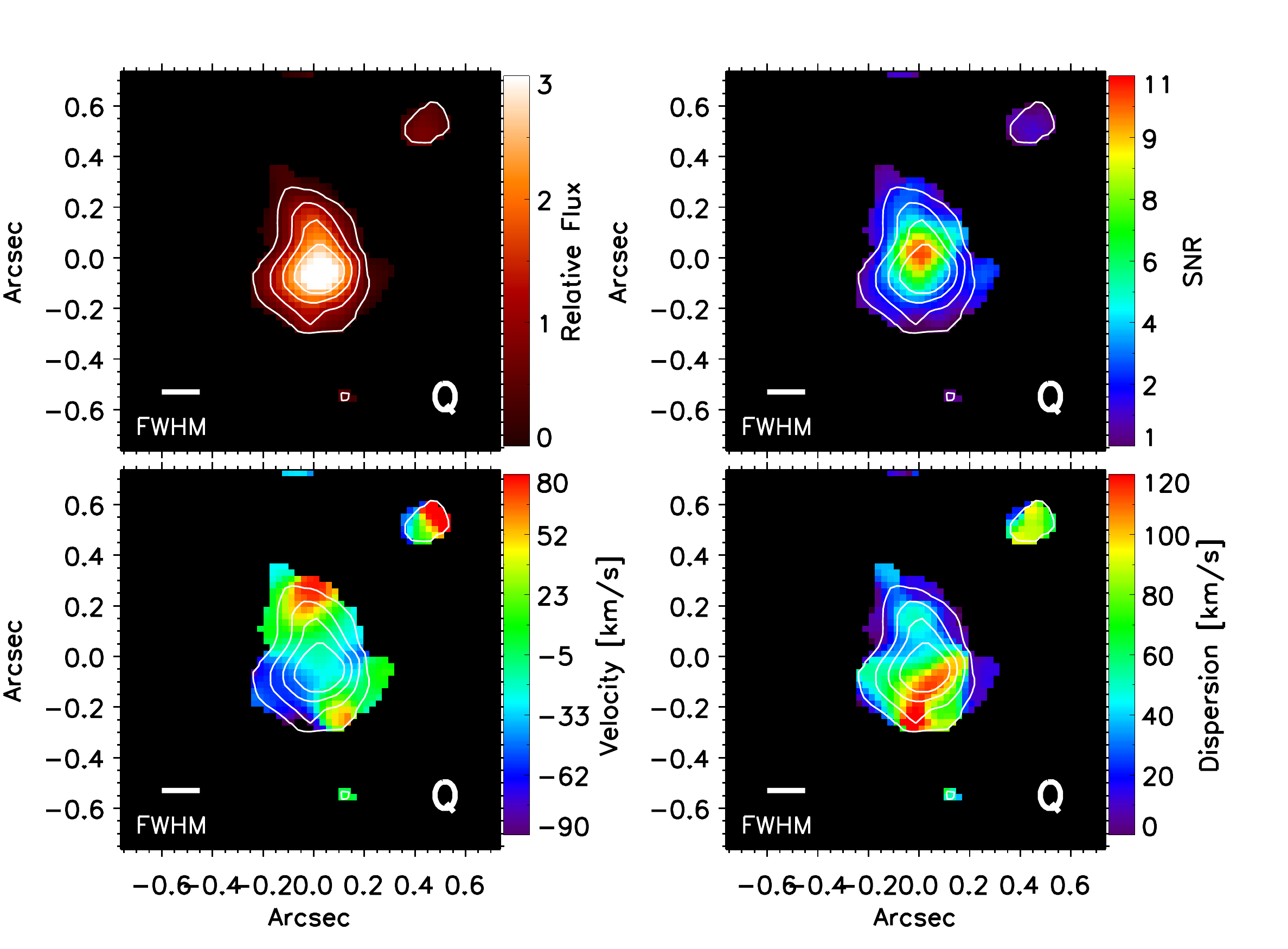}
\caption{ \ha\ intensity (top left), velocity (bottom left), velocity dispersion (bottom right) and  S/N (top right) maps.  The orientation is the standard North, up and East, to the left. The velocity is relative to $z =$\hazabs , the best-fit redshift determined from the composite spectrum shown in Figure~\ref{fig:halpha}.  Individual spaxels are 0.025 arcsec$^2$.  The FWHM =$\sim$ 0.15\arcsec\ of the PSF after smoothing is shown in the lower left corner of the intensity map.  At the redshift of the \dla\ 1\arcsec\ corresponds to $\sim$8.3 kpc.  The quasar, indicated by a 'Q', is located in the lower right hand corner at approximately x = 0.5\arcsec\ and y = $-$0.6\arcsec .  
}
\label{fig:havelocity}
\end{figure*}

\begin{figure*}
\includegraphics[scale=0.78]{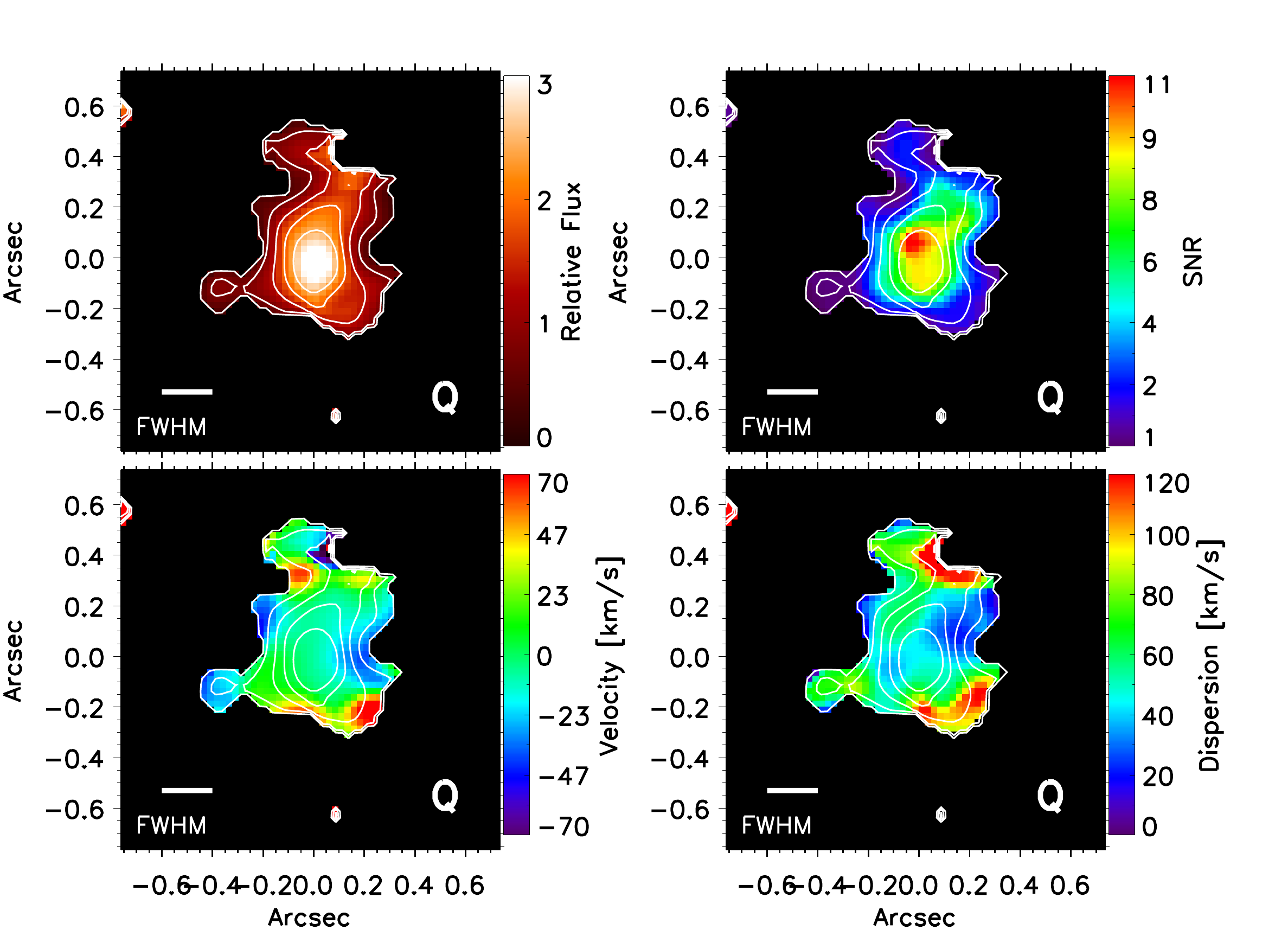}
\caption{ \oiii\ $\lambda$5007 intensity (top left), velocity (bottom left), velocity dispersion (bottom right) and  S/N (top right) maps.   The orientation is the standard North, up and East, to the left.  The velocity is relative to $z =$\oiiiazabs , the best-fit redshift determined from the composite spectrum shown in Figure~\ref{fig:oiii}.  Individual spaxels are 0.025 arcsec$^2$. The FWHM =$\sim$ 0.20\arcsec\ of the PSF after smoothing is shown in the lower left corner of the intensity map.  At the redshift of the \dla\ 1\arcsec\ corresponds to $\sim$8.3 kpc.  The quasar, indicated by a 'Q', is located in the lower right hand corner at approximately x = 0.5\arcsec\ and y = $-$0.6\arcsec .  
}
\label{fig:oiiivelocity}
\end{figure*}

 We present the \ha\ and \oiii\ emission maps in Figures~\ref{fig:havelocity} and ~\ref{fig:oiiivelocity}, respectively.  Each figure contains a relative flux map, centered on the peak emission location (top left), and the corresponding velocity map (bottom left), and velocity dispersion map (bottom right). The average FWHM of the point-spread function (PSF) as measured by the image of the quasar (not pictured) is indicated by the white bar, and is FWHM $\sim$0.15\arcsec\ and FWHM $\sim$ 0.20\arcsec\ for the \ha\ and \oiii\ maps, respectively. In all maps the quasar (not pictured) is located at x$\sim$0.5\arcsec\ and y$\sim -$0.6\arcsec , and indicated by a `Q.'  The distance from the quasar to the center of the \dla\ emission is $\sim$0.7\arcsec , corresponding to a physical distance at the redshift of the \dla\ of $\sim$5.8 kpc (where 1\arcsec\ corresponds to $\sim$8.34 kpc).   
 
In order to gauge the significance of the detected line emission, we calculate the S/N of the line detection in each spaxel by measuring the standard deviation of the noise in a nearby spectral region free from skylines or other emission-lines.  
We then estimate the  S/N as the ratio of the amplitude of the best-fit Gaussian to the standard deviation of the noise region (where the amplitude of the best-fit Gaussian is taken to be the maximum of the Gaussian fit minus the continuum level of the fit). Variations in the S/N from spaxel to spaxel are shown in the upper right panel of Figures ~\ref{fig:havelocity} and ~\ref{fig:oiiivelocity}.

   \section{Results}~\label{sec:results}
   
   In this section we discuss the estimates of mass and star formation rate surface density as well as the implications of the derived kinematics of the galaxy.

   \subsection{Dynamical mass estimate}
   We estimate the dynamical mass of the galaxy within the radius probed by the \ha\ emission using equation 2 from ~\cite{law09},
   
   \begin{equation}
   M_{dyn} = \frac{C \sigma ^{2}_{net} r}{G}
   \end{equation}
   
 \noindent where C = 5 for a uniform sphere ~\citep{erb06}, $\sigma _{net}$ $\sim$ \hasig\ \kms  , as measured from the stacked \ha\ spectrum shown in Figure~\ref{fig:halpha}, and the radius $r$ is approximated to be 0.25 arcsec, which corresponds to $\sim$2.1 kpc at the redshift of the galaxy.  The calculated dynamical mass is therefore, $M_{dyn}$ = \hamass\ M$_{\odot }$.  We can compare this with the dynamical mass derived by ~\cite{krogager13} of $M_{dyn}$ = 2.5 $\times$ 10$^9$ M$_{\odot }$.  We note that while ~\cite{krogager13} report a $\sigma$ value nearly equal to the value reported here  ($\sigma$ $\sim$ 49.1 \kms ), they measure a smaller galactic size, with semi-major axis a$_{e}$ = 1.12 kpc from their HST UV imaging, and as a result, estimate a smaller $M_{dyn}$.  

We find that the dynamical mass of DLA 2222$-$09 is slightly larger than that of the Small Magellanic Cloud (SMC), measured to be $M_{dyn}$ = 2.4 $\times$ 10$^9$ M$_{\odot }$. Moreover, it is similar to the low end of the dynamical mass range of the high redshift star forming galaxies studied by ~\cite{law09}, which range from $M_{dyn}$ = 3 $\times$ 10$^9$ M$_{\odot }$  to $M_{dyn}$ = 25 $\times$ 10$^9$ M$_{\odot }$.
 
Given the observational limitations, there are currently few direct measurements of dynamical masses of \dlas\ reported in the literature.  ~\cite{chengalur02} use \hi\ 21cm imaging to estimate the  dynamical mass of a low redshift \dla , at $z = 0.009$, to be $M_{dyn}$ = 5 $\times$ 10$^9$ M$_{\odot }$, similar to the galaxy presented here.  While ~\cite{peroux11} used the VLT/SINFONI IFU to map \ha\ emission and estimate the dynamical masses of a DLA and sub-DLA at $z\sim$1 to be significantly larger at $M_{dyn} = 2.0 \times 10^{10}$ M$_{\odot }$ and $M_{dyn} = 7.9 \times 10^{10}$ M$_{\odot }$, respectively.

\subsection{Star formation rate surface density and gas mass estimates}

In order to use the Kennicutt-Schmidt relation ~\citep{kennicutt98schmidt} to estimate the gas mass of the galaxy, we first estimate the star formation rate surface density, $\Sigma _{SFR}$.  We do this by calculating the star formation rate, as in equation~\ref{eqn:sfr}, in each 0.025 arcsec$^2$ spaxel, and use the scale 1 arcsec = 8.338 kpc at the redshift of \dla\ 2222$-$09 (note: spaxel size is smaller because of oversampling).  
The $\Sigma _{SFR}$ peaks at the center of the \ha\ emission, to a maximum value of $\Sigma _{SFR}$ = \sigsfrpeak\ M$_{\odot}$ yr$^{-1}$ kpc$^{-2}$. We find a mean star formation rate surface density of $\langle \Sigma _{SFR} \rangle$ = \sigsfr\ M$_{\odot}$ yr$^{-1}$ kpc$^{-2}$.  This value is about an order of magnitude lower than the $\Sigma _{SFR}$ found for Lyman Break Galaxies (LBGs) of ~\cite{law07}.   According to the results of ~\cite{kennicutt98schmidt}, this rate places DLA2222$-$09 at the high end of the sample of normal disk galaxies and at the the low end of starburst samples.  

We use the Kennicutt-Schmidt relation ~\citep{kennicutt98schmidt},

\begin{equation}
\Sigma _{SFR} = (2.5 \pm 0.7) \times 10^{-4}  \left(\frac{\Sigma_{gas}}{1 M_{\odot} \mathrm{pc}^{-2}}\right)^{1.4 \pm 0.15} \cmma\
\end{equation}

\noindent to calculate the gas mass surface density,  $\Sigma _{gas}$.  We find a mean gas mass surface density of  $\langle \Sigma _{gas} \rangle$ = \siggas\ M$_{\odot}$ pc$^{-2}$.  Integrating over the extent of the emission region, we find a total gas mass of M$_{gas}$ = \totgas\ M$_{\odot}$ .  Considering the uncertainties and large inherent errors, our results agree fairly well with those of \cite{krogager13} who estimate the gas mass of this galaxy to be M$_{gas}$ = 1 $\times$ 10$^9$ M$_{\odot}$.
 We find that this gas mass is approximately an order of magnitude less than those of star forming galaxies at $z\sim 2$ studied by ~\cite{erb06}, who find a mean inferred gas mass is $\langle\ M_{gas} \rangle\ = (2.1 \pm\ 0.1) \times 10^{10} M_{\odot}$.  
 We estimate the gas fraction of \mydla , calculated as $f_{gas} = \frac{M_{gas}}{M_{gas} + M_{dyn}}$, to be $f_{gas} \sim$40\%, in agreement with the ~\cite{krogager13} estimate.
 
\begin{figure*}
\plotone{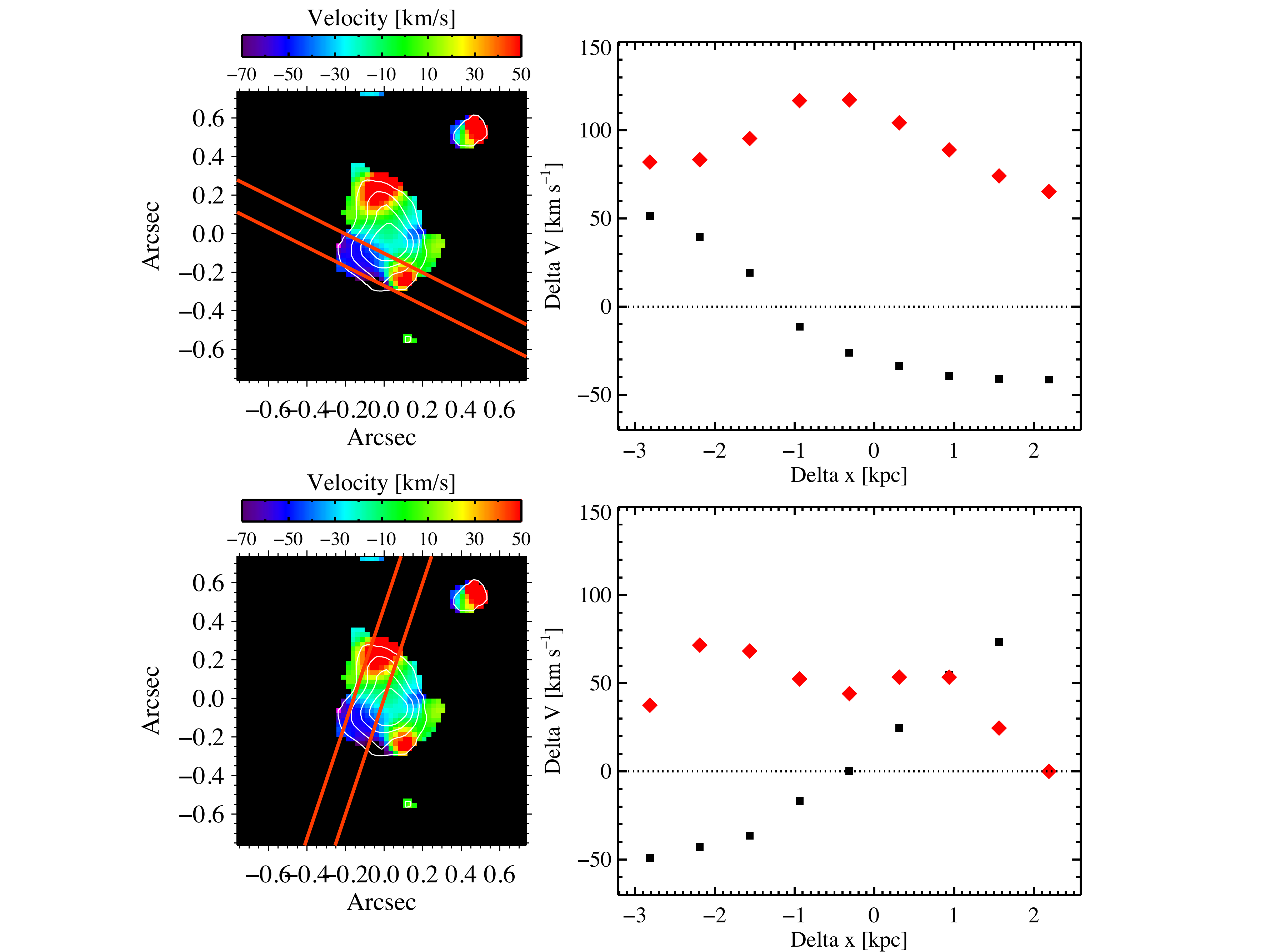}
\caption{ Laying down artificial slits on the \ha\ emission map.  Relative velocity (black squares) and velocity dispersion (red diamonds) extracted from the slit are plotted on the right, where the x-axis is position along the slit in kpc, while the y-axis is the velocity difference from the central redshift.
}
\label{fig:slitsummary}
\end{figure*}

 \input{summaryresults.tex}

 \subsection{Morphology and kinematics}

A comparison of the \ha\ and \nii\ emission maps, shown in Figures~\ref{fig:havelocity} and ~\ref{fig:oiiivelocity}, reveals much information about the morphological and kinematical state of \dla\ 2222$-$09.  First, we find that while the peak \ha\ and \oiii\ emission correspond spatially, the \oiii\ emission is stronger and slightly more spatially extended than that of the H$\alpha$.  In both cases, the roughly circular morphology makes it difficult to determine a potential kinematic major axis, and suggests a face-on disk orientation.  
The velocity profiles, shown in the bottom left of Figures~\ref{fig:havelocity} and ~\ref{fig:oiiivelocity}, reinforce this picture, as there is no clear kinematic signature of rotation. While we do detect some `clumps' of red and blue shifted emission, the velocity across the central emission region, as seen in both the \ha\ and \oiii\ velocity maps, appears relatively constant. 

In the most simplistic approximation we can estimate the amount of dynamical support provided by rotation by calculating $\frac{v_{shear}}{\sigma_{mean}  }$, where the velocity shear, v$_{shear} = \frac{1}{2} (v_{max} - v_{min}$), and $\sigma _{mean}$ is the mean velocity dispersion of the system.  We find $v_{shear}$ $\sim$ 119  km s$^{-1}$, and $\sigma _{mean} \sim$ 57  km s$^{-1}$, resulting in $\frac{v_{shear}}{\sigma_{mean}  } \sim$ 2.1.  However, we caution that while this seems to be indicative of a rotationally supported disk (i.e. ~\cite{forster09, law09}), these values were derived by averaging over the system as a whole, and we do not see evidence of the classical, edge-on disk rotation in which the velocity dispersion should peak at the center of rotation.  

To quantitatively search for signs of edge-on disk rotation, we created artificial slits and extracted the velocity and velocity dispersion, $\sigma $. The slit widths were made to match the FWHM of the seeing, 0.15\arcsec . We then placed the slit on the region of interest and extracted the median velocity and $\sigma $ over dispersion regions equal to 1/2 of the FWHM, or 0.075\arcsec . For example, in Figure~\ref{fig:slitsummary}, top, we place the slit at position angle (PA) $\sim$70$^{\circ}$ east of north, and find that the velocity and velocity dispersion are indicative of an edge-on disk rotation pattern in which the velocity smoothly varies from $+$50 \kms to $-$50 \kms across the span of $\sim$5 kpc while the corresponding velocity dispersion peaks in the middle at $v = 0$ \kms relative velocity and $\sigma \sim$120   km s$^{-1}$.  However, despite the intriguing kinematic possibility, we doubt the interpretation of this as disk rotation because 1) the center of rotation is not aligned with the peak of \ha\ emission, as would be expected, and 2) the large $\sigma$ and relatively small velocity shear would indicate a dispersion dominated system rather than a large, rotationally supported disk. In Figure~\ref{fig:slitsummary}, bottom, we align the slit at PA$\sim$15$^{\circ}$ west of north, to demonstrate how a velocity profile that seems indicative of rotation, is clearly not when the associated velocity dispersion is examined.  

While there is no clear evidence of disk-like morphology or rotation, we note that the kinematic signatures that are seen, agree, at least qualitatively, across the independent \ha\ and \oiii\ emission-line maps.  For example, both the \ha\ and \oiii\ velocity maps show regions of redshifted emission both north and south of the central emission peak, while blueshifted regions appear along a roughly east-west axis.  
Interestingly, these blueshifted regions appear, at least qualitatively, to align with the major axis of rest-frame UV emission detected by ~\cite{krogager13}.  
From their {\it HST} F606W image, ~\cite{krogager13} find that \mydla\ has a compact, elongated structure indicative of an edge-on disk.   Given that both \ha\ and the rest-frame UV should trace the sites of active star formation, it is not clear why the galaxy morphology implied by the {\it HST} image and the \ha\ emission map presented here, should be different.  
One explanation may be that foreground dust, perhaps associated with the blueshifted \ha\ emission, is obscuring background UV emission, causing the appearance of an elongated, disk-like morphology in the rest-frame UV.  However, given the generally low S/N of these blue and redshifted regions (S/N $\lesssim$ 2), we are hesitant to over-interpret their significance or meaning.

\subsection{Evidence of metallicity gradient or metal-enriched winds/outflow/infall?}~\label{sec:nii}
The relatively high metallicities of \dlas\ compared with the Lyman$-\alpha$ forest ~\citep{schaye03, aguirre04}, in addition to the Hubble Deep Field constraint on {\it in situ} star formation ~\citep{wolfe06}, suggest that metal enrichment of \dlas\ is at least partially due to a secondary process such as the ejection and infall of metal-enriched winds.  The observations of \mydla\ presented here may support just such a scenario.  The emission-line metallicity measured at the center of \mydla , \niimetsolaroptpercent\ solar, is more than twice as large as the metallicity determined from the absorption-lines, $\sim$30\% solar, measured $\sim$ 6 kpc away in front of the background quasar.  While the errors inherent in the emission-line derived metallicity via the $N2$ relation are large and these metallicities could be consistent with one another, we note that this could also be evidence of either a metallicity gradient in the galaxy, or the presence of metal-enriched winds escaping (or falling back onto) the galaxy.

\subsection{Comparison with Keck/HIRES absorption-line data}~\label{sec:abs}

In this section we compare the OSIRIS data with a Keck/HIRES echelle spectrum of \mydla\ obtained from the Keck Observatory Archive.  The spectrum was taken on August 17, 2006 with the C1 decker for a spectral resolution of $\sim$6   km s$^{-1}$.  The final spectrum, a coadd of 2 5400s exposures for a total exposure time of 10800s, has a median S/N of $\sim$16 pixel$^{-1}$.

As reported in ~\cite{jorgenson13a}, the metallicity as measured from the HIRES spectrum is [M/H] = $-$0.56 $\pm$0.10, which corresponds to a metallicity of $\sim$30\% solar.  The velocity interval containing 90 per cent of the integrated optical depth of the low-ion metallic gas ~\citep{pro97} is also relatively large at $\Delta v_{90}$ = 179   km s$^{-1}$.  Similarly, the equivalent width of the Si\, II $\lambda$1526 line, thought to be a proxy for mass ~\citep{pro08}, is on the high end of \dla\ samples at W$_{1526}$ = 1.23\AA ~\citep{jorgenson13a}. Unfortunately, it is not possible to measure the level of star formation activity via the \ciistr\ technique ~\citep{wolfe03a}, as the \ciistr $\lambda$1335.7 absorption-line is blended with the strong CII$\lambda$ 1334 line.  However, all previously mentioned quantities, including metallicity, low-ion velocity width and the Si\, II $\lambda$1526 equivalent width, indicate that \mydla\ is a typical `high$-$cool' \dla\ in which one might expect heating to be derived primarily from a nearby LBG galaxy ~\citep{wolfe08}.

\begin{figure}
\plotone{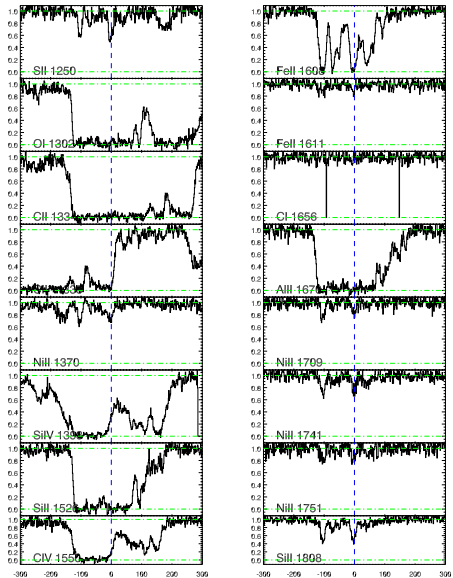}
\caption{ Select low- and high-ion velocity transitions from the Keck/HIRES spectrum.  The velocity $v = 0$ \kms\ shown here corresponds to the fiducial redshift, taken to be $z_{HIRES}$ = \hireszabs .  For comparison, on this velocity scale the peak of \ha\ emission falls at \havel\   km s$^{-1}$. 
}
\label{fig:hires}
\end{figure}

In Figure~\ref{fig:hires}, we present a selection of low- and high-ion velocity profiles of \mydla .  We arbitrarily define the fiducial absorption redshift of the system, $z_{HIRES} =$ \hireszabs , to be located at the central, low-ion velocity component of highest optical depth, denoted by $v = 0$   km s$^{-1}$.  As seen from the unsaturated species, e.g. Si\,II $\lambda$1808 and Si\,II $\lambda$1250, there is a second velocity component of high optical depth located at $v \sim -$100   km s$^{-1}$.  Interestingly, the central redshift of the emission-lines, both \ha\ and \oiii , places them at $v \sim -$50  km s$^{-1}$, in the middle of the two strongest low-ion velocity components.  This remarkable coincidence between the emission-line redshift and the absorption-line redshift, measured $\sim$6 kpc away, was noted by ~\cite{fynbo10}, who suggest it could be explained by the quasar line of sight passing parallel to the rotation axis of a face-on disk.
In addition, we see that the high-ionization lines, such as C\,IV $\lambda$1550 and Si\,IV $\lambda$1393, have velocity profiles nearly identical to those of the low ions, with just slightly stronger redshifted velocity components.  This is unusual for \dla\ velocity profiles, in which the high-ions typically contain velocity components that are offset and extended in comparison with the low-ionization lines, and as a result thought to be tracing winds/outflows.

\begin{figure*}
\includegraphics[scale=0.5]{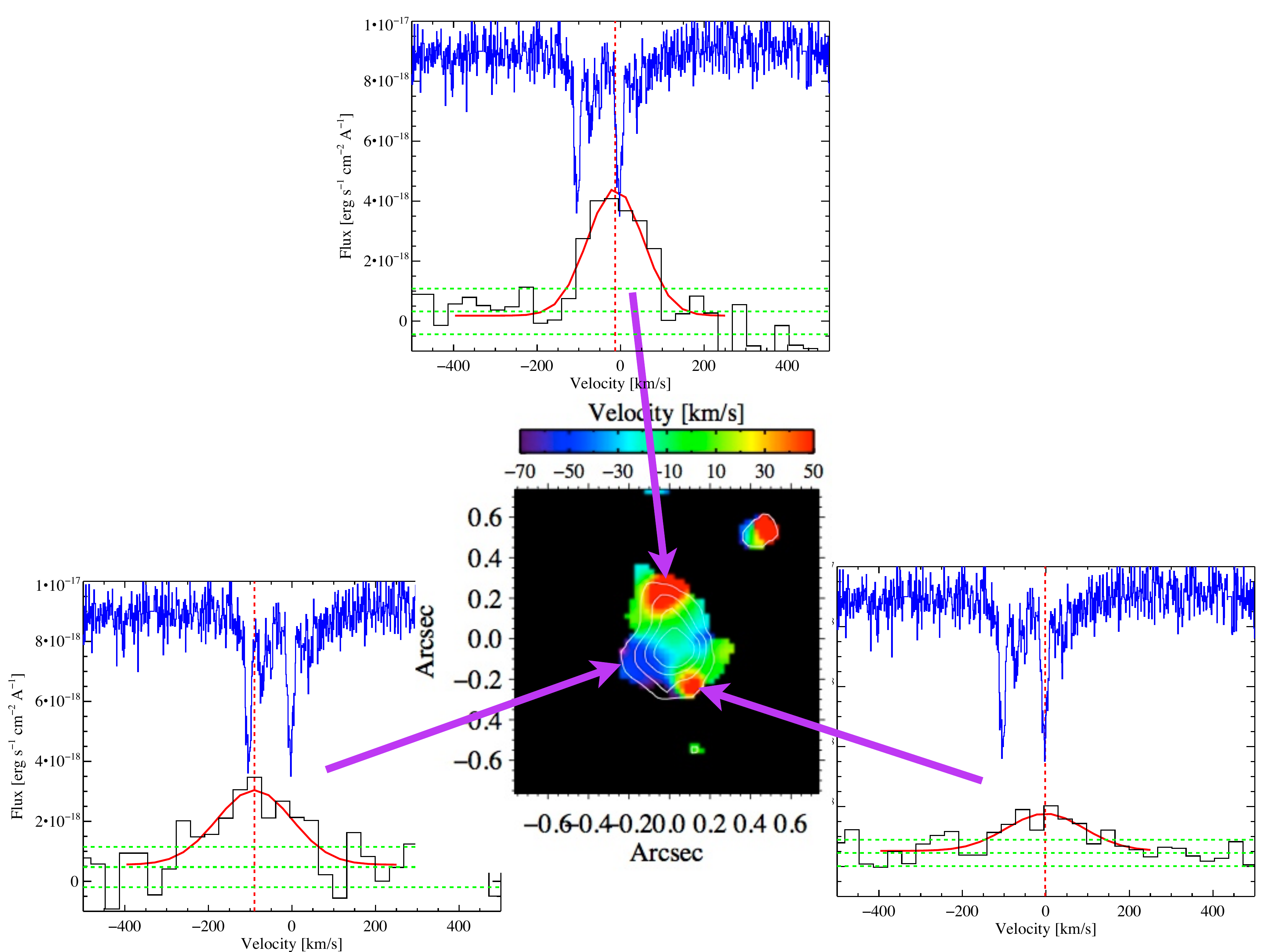}
\caption{ \ha\ emission profiles over small spatial regions compared with the Keck/HIRES spectrum of SiII$\lambda$1808 line (overplotted in blue).  It is seen that the blue-shifted \ha\ region (lower left) corresponds in velocity space to the blue low-ion velocity component seen in the Keck/HIRES spectrum (here, at $v\sim $100\kms ).  Similarly, the redshifted \ha\ regions (top and lower right) closely correspond in velocity space to the red low-ion velocity component, here at $v\sim 0$  km s$^{-1}$.  
}
\label{fig:velcomparehires}
\end{figure*}

While it may be purely coincidental given the 6 kpc spatial separation, we note that the red and blue shifted emission clumps are shifted to velocities corresponding to the two highest optical depth low-ion absorption lines at $v \sim 0$ and $v \sim -$100 km s$^{-1}$.  In Figure~\ref{fig:velcomparehires} we compare sub-stacks of \ha\ emission from spatial regions that appeared either blue or redshifted with respect to the central emission velocity, with the velocity profile of the unsaturated Si\,II $\lambda$1808 absorption feature.  There exists a remarkable agreement between the redshifts (or velocities) of the emission sub-clumps and the two strongest low-ion absorption components.  In the top and lower right panel, the best-fit redshift of the sub-stack \ha\ emission corresponds almost exactly with that of the Si\,II $\lambda$1808 velocity component at $v = 0$  km s$^{-1}$, while in the lower left panel, the best-fit \ha\ emission-line corresponds to within the errors with the Si\,II $\lambda$1808 velocity component at $v \sim -100$ km s$^{-1}$.

\subsubsection{Molecular Hydrogen}
Given the high metallicity and relatively large star formation rate of \mydla , one might expect to detect a large amount of molecular gas.  However, this gas is not detected in the HIRES spectrum.  Using the HIRES spectrum, ~\cite{jorgenson13b} place an upper limit on the amount of molecular hydrogen (\htwo ) in the system, of N(\htwo ) $\lesssim 2 \times 10^{14}$ cm$^{-2}$.  This corresponds to the low molecular fraction of $f \lesssim 10^{-5.8}$.  In addition, there is no evidence of neutral carbon (\ci ) absorption, a species that is often associated in \dlas\ with the cold, dense gas required for the presence of \htwo ~\citep{jorgenson10}.  While it is clear that there is not much, if any, \htwo\ in this \dla , the \ha\ emission indicates a SFR of $\sim$ 10 M$_{\odot}$ yr$^{-1}$.  Presumably, \htwo\ had to exist at some point in order to create the stars producing the \ha\ emission.  Is it merely chance that the sight line probed by the background quasar is devoid of molecules?  Or is the molecular gas more centrally concentrated resulting in the apparently low \htwo\ covering factor found by surveys for \htwo\ ~\citep{jorgenson13b}?  Or could this be an example of a metal-rich, star-forming \dla\ galaxy in which copious star formation has temporarily depleted the supply of molecular gas?  Followup observations to search for molecular emission might help answer these questions.  

\subsubsection{Lyman$-\alpha$ emission}
Strong asymmetric Lyman$-\alpha$ emission in the trough of \mydla\ is detected by ~\cite{fynbo10} and ~\cite{krogager13}, with a flux of F(Ly$\alpha$) = (14.3 $\pm$ 0.3) $\times$ 10$^{-17}$ ergs s$^{-1}$ cm$^{-2}$.  However, we do not detect Lyman$-\alpha$ emission in the HIRES spectrum.  This is likely due to an unfortunate positioning of the slit during the Keck/HIRES observations. However,  we do find evidence of a slight rise in the zero level on the red side of the \dla\ trough that is consistent with the Lyman-$\alpha$ emission profile as shown in ~\cite{krogager13}.

\section{Summary}~\label{sec:summary}   
We present the first Keck/OSIRIS IFU observations of a high redshift \dla\ galaxy that, aided by LGSAO, spatially resolve the \ha\ and \oiii\ emission.  With a star formation rate of nearly 10 M$_{\odot}$ yr$^{-1}$, and a dynamical mass, M$_{dyn}$ = \hamass\ M$_{\odot}$, \mydla\ appears similar to the low-mass end of the high redshift star forming galaxies studied by ~\cite{law09}.  We detect \nii\ emission with 2.5$\sigma$ significance and estimate a metallicity of \niimetsolaroptpercent\ solar in the central galactic region.  When compared with the absorption-line metallicity,  $\sim$30\% solar, measured 6 kpc away, this may suggest either a metallicity gradient or the presence of metal enriched out/inflow.  

Kinematically, we find the central emission regions of maximum flux and S/N to be relatively constant, showing no evidence of a smoothly varying velocity gradient consistent with the rotation of a disk viewed edge-on, as suggested by the HST rest-frame UV images of ~\cite{krogager13}.  We do detect several red and blueshifted `clumps' of emission which could be analogous to the kiloparsec-sized clumps commonly seen in high redshift star forming galaxies, e.g ~\cite{elmegreen09a, elmegreen09b}.  The lack of evidence of ordered rotation, in addition to the generally circular morphology indicated by the emission lines and the remarkable coincidence of emission and absorption-line redshifts, support the interpretation, originally proposed by ~\cite{fynbo10}, that \mydla\ is a disk viewed nearly face-on.

The observations presented here highlight the potential for using LGSAO+IFU instruments on 10-meter class telescopes to finally achieve the long-sought goal of imaging the host galaxies of \dlas .  
We have demonstrated that it is possible, with reasonable exposure times of a few hours per object, to detect and map the relatively faint \dla\ emission around a bright, central quasar.  Only by now increasing the sample of mapped \dla\ emitters will we finally be able to craft a better understanding of the nature of these elusive \dla\ systems and their role in galaxy formation and evolution.
   
\acknowledgments
We would like to thank David Law, Tiantian Yuan, Randy Cambell, Jim Lyke, and Jessica Lu for many beneficial discussions, John O'Meara for providing the HIRES spectrum and Jeff Cooke for providing comments on an early draft.  R. A. J. gratefully acknowledges support from the NSF Astronomy and Astrophysics Postdoctoral Fellowship under award AST-1102683.  
The authors wish to recognize and acknowledge the very significant cultural role and reverence that the summit of Mauna Kea has always had within the indigenous Hawaiian community.  We are most fortunate to have the opportunity to conduct observations from this mountain.

\bibliographystyle{apj}
\bibliography{regina.bib}

 \end{document}

%% file: xavier_defs.tex
\def\aap{A \& A}

\def\apjl{ApJL}
\def\apjs{ApJS}

\def\lya{Ly$\alpha$ }
\def\ha{H$\alpha$ }

\def\kms{km~s$^{-1}$ }

\def\s-1{s$^{-1}$}
\def\Hz-1{Hz$^{-1}$}

\def\sci#1{{\; \times \; 10^{#1}}}

\def\l1l2{\lambda_2 \; {\rm and} \; \lambda_1}

\def\intl{\int\limits}

\def\eht1{{\rm \hat e_1}}

\def\ltk{\left [ \,}
\def\ltp{\left ( \,}
\def\ltb{\left \{ \,}
\def\rtk{\, \right  ] }
\def\rtp{\, \right  ) }
\def\rtb{\, \right \} }

\def\d3x{d^3 x}

\def\cmma{\;\;\; ,}
\def\perd{\;\;\; .}

\def\cm#1{\, {\rm cm^{#1}}}

%% file: compare_results.tex
\begin{center}
\begin{deluxetable*}{cccc}
\tablewidth{0pc}
\tablecaption{Summary of $H\alpha $ Flux, Luminosity and SFR for DLA 2222$-$09 \label{tab:results}}
\tabletypesize{\scriptsize}
\tablehead{\colhead{Author} &
\colhead{F($H\alpha$ )} &
\colhead{L$(H\alpha$ )} &
\colhead{SFR($H\alpha$ )} \\
 & \colhead{[ergs s$^{-1}$ cm$^{-2}$]} & \colhead{[ergs s$^{-1}$]}  & \colhead{[M$_{\odot} \ yr^{-1}$]} }
\startdata
This work & \haflux\  & \halum\  & \hasfr\  \\
F10$^{a}$ & 2.5 $\times$ 10$^{-17}$ & 1.1 $\times$ 10$^{42}$ & 10 \\
P12$^{b}$ & (8.7 $\pm$ 2.6) $\times$ 10$^{-17}$ & (3.85 $\pm$ 0.11) $\times$ 10$^{42}$ & 17.1 $\pm$ 5.1 \\
P13$^{c}$ & 5.6 $\times$ 10$^{-17}$ & - & - \\
K13$^{d}$ & (5.7 $\pm$ 0.3) $\times$ 10$^{-17}$ & (2.4 $\pm$ 0.10) $\times$ 10$^{42}$ & 12.7$^{e}$ $\pm$ 0.7 \\
\enddata
\tablenotetext{a}{~\cite{fynbo10}}
\tablenotetext{b}{~\cite{peroux12}}
\tablenotetext{c}{~\cite{peroux13}, Luminosity and SFR not provided.}
\tablenotetext{d}{~\cite{krogager13}}
\tablenotetext{e}{Using dust extinctiion derived from their UV data.}
\end{deluxetable*}
\end{center}

%% file: summarylines.tex
\LongTables
\begin{center}
\begin{deluxetable}{llc}
\tablewidth{0pc}
\tablecaption{Line Diagnostics of DLA 2222$-$09 \label{tab:summarylines}}
\tabletypesize{\scriptsize}
\tablehead{\colhead{Quantity} & \colhead{Units} &
\colhead{ Measured}} \\
\startdata
$z_{HIRES}$&& \hireszabs\ \\
\hline
$z$(H$\alpha$) && \hazabs\ \\
$\Delta v$(H$\alpha$)$^{a}$ & [km s$^{-1}$] & \havel\ \\
FWHM(H$\alpha$) & [km s$^{-1}$] & \hafwhm\ \\
FWHM(H$\alpha$)$^{b}$ & [km s$^{-1}$] &\hafwhminit\ \\
\hline
$z$(\nii ) && \niizabsopt\ \\
$\Delta v$(\nii )$^{a}$ & [km s$^{-1}$] & \niivelopt\ \\
FWHM(\nii ) & [km s$^{-1}$] &\niifwhmopt\ \\
FWHM(\nii )$^{b}$ & [km s$^{-1}$] &\niifwhmoptinit\ \\
\hline
$z$(\oiii\ $\lambda$5007) && \oiiiazabs\ \\
$\Delta v$(\oiii$\lambda$5007 )$^{a}$ & [km s$^{-1}$] & \oiiiavel\ \\
FWHM(\oiii\ $\lambda$5007) & [km s$^{-1}$] & \oiiiafwhm\ \\
FWHM(\oiii\ $\lambda$5007)$^{b}$ & [km s$^{-1}$] &\oiiiafwhminit\ \\
\hline
$z$(\oiii\ $\lambda$4959) && \oiiibzabs\ \\
$\Delta v$(\oiii\ $\lambda$4959)$^{a}$ & [km s$^{-1}$] & \oiiibvel\ \\
FWHM(\oiii\ $\lambda$4959) & [km s$^{-1}$] & \oiiibfwhm\ \\
FWHM(\oiii\ $\lambda$4959)$^{b}$ & [km s$^{-1}$] &\oiiibfwhminit\ \\
\enddata
\tablenotetext{a}{Velocity difference between $z_{HIRES}$ and the given transition.}
\tablenotetext{b}{Effects of instrumental smoothing taken out in quadrature.}
\end{deluxetable}
\end{center}

%% file: summaryresults.tex
\LongTables
\begin{center}
\begin{deluxetable}{llc}
\tablewidth{0pc}
\tablecaption{Summary of Results for DLA 2222$-$09 \label{tab:summaryresults}}
\tabletypesize{\scriptsize}
\tablehead{\colhead{Quantity} & \colhead{Units} &
\colhead{ Measured}} \\
\startdata
F(H$\alpha$ )$^a$ & [ergs s$^{-1}$ cm$^{-2}$] & \haflux\   \\
L(H$\alpha$ ) & [ergs s$^{-1}$]   & \halum\ \\
M$_{dyn}$ & [M$_{\odot}$] & \hamass\ \\
SFR & [M$_{\odot}$yr$^{-1}$] & \hasfr  \\
$\langle\ \Sigma _{SFR} \rangle $ & [M$_{\odot}$ yr$^{-1}$ kpc$^{-2}$] & \sigsfr\ \\
Peak $\Sigma _{SFR}$ & [M$_{\odot}$ yr$^{-1}$ kpc$^{-2}$] & \sigsfrpeak\ \\
$\langle\ \Sigma _{gas} \rangle $ & [M$_{\odot}$ pc$^{-2}$] & \siggas\ \\
M$_{gas}$ & [M$_{\odot}$]& \totgas\ \\
$f_{gas}$ & &$\sim 40\%$ \\
F(\nii )$^a$ & [ergs s$^{-1}$ cm$^{-2}$] & \niifluxopt\   \\
L(\nii ) & [ergs s$^{-1}$]   & \niilumopt\ \\
12+log(O/H) & & \niimetNopt\ \\
metallicity & \% of solar &  \niimetsolaropt\ \\
F(\oiii\ $\lambda$5007)$^a$ & [ergs s$^{-1}$ cm$^{-2}$] & \oiiiaflux\   \\
L(\oiii\ $\lambda$5007) & [ergs s$^{-1}$]   & \oiiialum\ \\
F(\oiii\ $\lambda$4959)$^a$ & [ergs s$^{-1}$ cm$^{-2}$] & \oiiibflux\   \\
L(\oiii\ $\lambda$4959) & [ergs s$^{-1}$]   & \oiiiblum\ \\
\enddata
\tablenotetext{a}{1$\sigma$ uncertainties are determined after subtraction of the gaussian model.}
\end{deluxetable}
\end{center}